\documentclass{article}
\usepackage{nips10submit_e,times}
\usepackage{graphicx}
\usepackage{amsmath,amsfonts,amssymb,bbm}
\usepackage{graphicx,subfigure}
\usepackage{epsfig,psfrag,color,url}
\usepackage{algorithm,algorithmic}
\usepackage{textcomp}
\usepackage{hyperref}

\title{Infinite Multiple Membership Relational Modeling for Complex Networks}

\author{
Morten M{\o}rup, Mikkel N. Schmidt and Lars Kai Hansen\\
Cognitive Systems Group\\
Technical University of Denmark\\
\texttt{\{mm,mns,lkh\}@imm.dtu.dk} }

\newcommand{\m}[1]{\boldsymbol{#1}}

\newcommand{\scprod}[1]{\displaystyle\mathop{\textstyle\prod}_{\makebox[0pt][c]{$\scriptstyle #1$}}}
\newcommand{\cprod}[1]{\prod_{\makebox[0pt][c]{$\scriptstyle #1$}}}
\newcommand{\csum}[1]{\sum_{\makebox[0pt][c]{$\scriptstyle #1$}}}

\makeatletter
\def\url@leostyle{%
  \@ifundefined{selectfont}{\def\UrlFont{\sf}}{\def\UrlFont{\footnotesize}}}
\makeatother
\begin{document}

\maketitle

\begin{abstract}
  Learning latent structure in complex networks has become an
  important problem fueled by many types of networked data originating
  from practically all fields of science. In this paper, we propose a
  new non-parametric Bayesian multiple-membership latent feature
  model for networks. Contrary to existing multiple-membership models
  that scale quadratically in the number of vertices the proposed
  model scales linearly in the number of links admitting multiple-membership analysis in large scale networks. We demonstrate a connection
  between the single membership relational model and multiple
  membership models and show on ``real'' size benchmark network data
  that accounting for multiple memberships improves the learning of
  latent structure as measured by link prediction while explicitly accounting for multiple membership result in a more compact representation of the latent structure of networks.
\end{abstract}

\section{Introduction}
The analysis of complex networks has become an important challenge
spurred by the many types of networked data arising in practically all
fields of science. These networks are very different in nature ranging
from biology networks such as protein interaction
\cite{Sun2003,Airoldi2008} and the connectome of neuronal connectivity
\cite{Watts1998} to the analysis of interaction between large groups
of agents in social and technology networks
\cite{Newman2001,Watts1998,Kemp2006,Xu2006}. Many of the networks exhibit a
strong degree of structure; thus, learning this structure facilitates both the understanding of network
dynamics, the identification of link density heterogeneities, as well
as the prediction of ``missing'' links.

We will represent a network as a graph $\mathcal{G}=(\mathcal{V},\mathcal{Y})$ where
$\mathcal{V}=\{v_1,\dots,v_N\}$ is the set of vertices and $\mathcal{Y}$ is the set of
observed links and non-links. Let $\m{Y}\in \{0,1,?\}^{N\times N}$ denote a link (adjacency) matrix where the element $y_{ij}=1$ if there
is a link between vertex $v_i$ and $v_j$, $y_{ij}=0$ if there is not a link, and $y_{ij}=?$ if the existence of a link is
unobserved. Furthermore, let $\mathcal{Y}_1$, $\mathcal{Y}_0$, and $\mathcal{Y}_?$ denote the set of links, non-links, and unobserved
links in the graph respectively.

Over the years, a multitude of methods for identifying latent
structure in graphs have been proposed, most of which are based on
grouping the vertices for the identification of homogeneous regions.
Traditionally, this has been based on various community detection
approaches where a community is defined as a densely connected subset
of vertices that is sparsely linked to the remaining network
\cite{Newman2004,Reichardt2006}. These structures have for instance
been identified by splitting the graph using spectral approaches,
analyzing flows, and through the analysis of the
Hamiltonian. Modularity optimization \cite{Newman2004} is a special
case that measures the deviation of the fraction of links within
communities from the expected fraction of such links based on their
degree distribution \cite{Newman2004,Reichardt2006}. A drawback,
however, for these types of analyses is that they are based on
heuristics and do not correspond to an underlying generative process.

\textbf{Probabilistic generative models: } Recently, generative models for complex networks have been proposed
where links are drawn according to conditionally independent Bernoulli
densities, such that the probability of observing a link $y_{ij}$ is given
by $\pi_{ij}$,
\begin{equation}
  \label{eq:py}
  p(\m{Y}|\m{\Pi}) = \prod_{(i,j)\in\mathcal{Y}}\pi_{ij}^{y_{ij}}(1-\pi_{ij})^{1-y_{ij}}.
\end{equation}
In the classical Erd\H{o}s-R\'enyi random graph model, each link is
included independently with equal probability $\pi_{ij}=\pi_0$;
however, more expressive models are needed in order to model complex
latent structure of graphs. In the following, we focus on two related
methods: latent class and latent feature models.

\textbf{Latent class models: }In latent class models, such as the stochastic block model
\cite{Nowicki2001}, also denoted the relational model (\textsc{rm}),
each vertex $v_i$ belongs to a class $c_i$, and the probability,
$\pi_{ij}$, of a link between $v_i$ and $v_j$ is determined by the
class assignments $c_i$ and $c_j$ as $\pi_{ij} = \rho_{c_ic_j}$. Here,
$\rho_{k\ell}\in[0,1]$ denotes the probability of generating a link
between a vertex in class $k$ and a vertex in class $\ell$. Inference
in latent class models involves determining the class assignments as
well as the class link probabilities. Based on this, communities can
be found as (groups of) classes with high internal and low external
link probability.

In the model proposed by \cite{Hofman2008} (\textsc{hw}) the class
link probability, $\rho_{k\ell}$, is specified by a within-class
probability $\eta_{c}$ and a between-class probability
$\eta_n$, i.e $\rho_{k\ell}=\eta_n(1-\delta_{k\ell})+\eta_c\delta_{k\ell}$.. Another intuitive representation, which we refer to as
\textsc{db}, is to have a shared between-class probability but allow
for individual within-class probabilities, i.e $\rho_{k\ell}=\eta_n(1-\delta_{k\ell})+\eta_k\delta_{k\ell}$.
Both of these representations are consistent with the notion of
communities with high internal and low external link density, and
restricting the number of interaction parameters can facilitate model
interpretation compared to the general \textsc{rm}.

Based on the Dirichlet process, \cite{Kemp2006,Xu2006} propose a
non-parametric generalization of the stochastic block model with a
potentially infinite number of classes denoted the infinite relational
model (\textsc{irm}) and infinite hidden relational model respectively. The latter generalizing the stochastic block model to simultaneously model potential vertex attributes. Inference in \textsc{irm} jointly determines the
number of latent classes as well as class assignments and class link
probabilities. This approach readily generalizes to the \textsc{hw}
and \textsc{db} parameterizations of $\boldsymbol{\rho}$.

\textbf{Latent feature models: }In latent feature models, the assumption that each vertex belongs to a
single class is relaxed. Instead it is assumed that each vertex $v_i$
has an associated feature $\m{z}_i$, and that probabilities of links
are determined based on interactions between features. This
generalizes the latent class models, which are the special case where
the features are binary vectors with exactly one non-zero element.
\begin{figure}[tbp]
  \centering
  \psfragscanon
  \psfrag{A}[bc][bc]{$v_1$}
  \psfrag{B}[bc][bc]{$v_2$}
  \psfrag{C}[bc][bc]{$v_3$}
  \psfrag{D}[bc][bc]{$v_4$}
  \psfrag{E}[bc][bc]{$v_5$}
  \psfrag{F}[bc][bc]{$v_6$}
  \psfrag{G}[bc][bc]{$v_7$}
  \psfrag{H}[bc][bc]{$v_8$}
  \psfrag{I}[bc][bc]{$v_9$}
  \psfrag{Y}[bc][bc]{$\m{Y}$}
  \includegraphics[width=70mm]{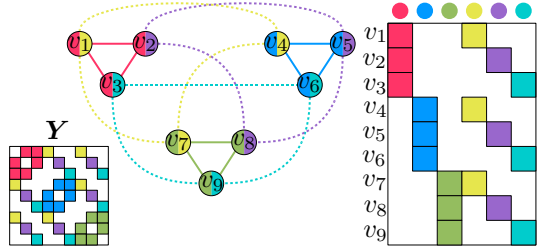}
 \caption{Left: Example of a simple graph where each of the vertices
   have multiple memberships indicated by colors. Right: The
   corresponding assignment matrix.}
  \label{fig:Clusters}
\end{figure}

Many latent feature models support the notion of discrete classes, but
allow for mixed or multiple memberships (see Figure~\ref{fig:Clusters}
for an illustration of a network with multiple class memberships). In
the mixed membership stochastic block model (\textsc{mmsb})
\cite{Airoldi2008} the vertices are allowed to have fractional class
memberships. In binary matrix factorization \cite{Meeds2007} multiple
memberships are explicitly modeled such that each vertex can be
assigned to multiple clusters by an infinite latent feature model
based on the Indian buffet process (\textsc{ibp})
\cite{Griffiths2006}. \cite{Miller2009} study this approach, for the
specific case of a Bernoulli likelihood, Eq.~(\ref{eq:py}), and extend
the method to include additional side information as covariates in
modeling the link probabilities. In their model, the probability of a
link $\pi_{ij}$ is specified by
$\pi_{ij}=f_{\sigma}\left(\sum_{k\ell}z_{ik}z_{j\ell}w_{k\ell}+s_{ij}\right)$,
where $f_{\sigma}(\cdot)$ is a function with range $[0,1]$ such as a
sigmoid, and $w_{k\ell}$ are weights that affects the probability of
generating a link between vertices in cluster $k$ and $\ell$. The term
$s_{ij}$ accounts for bias as well as additional side-information. For
example, if covariates $\m{\phi}_i$ are available for each vertex
$v_i$, \cite{Miller2009} suggest including the term
$s_{ij}=\beta
d(\m{\phi}_i,\m{\phi}_j)+\m{\beta}_i^\top\m{\phi}_i+\m{\beta}_j^\top\m{\phi}_j$,
where $\beta$, $\m{\beta}_i$, and $\m{\beta}_j$ are regression
parameters, and $d(\cdot,\cdot)$ is some possibly nonlinear function.

In general, the computational cost of the single membership clustering
methods mentioned above scales linearly in the number of links in the
graph. Unfortunately, existing multiple membership models
\cite{Airoldi2008,Meeds2007,Miller2009} scale quadratically in the
number of vertices, because they require explicit computations for all
links and non-links. This renders existing multiple membership
modeling approaches infeasible for large networks. Furthermore, determining the multiple membership
assignments is a combinatorial challenge as the number of potential
states grow as $2^{KN}$ rather than $K^N$ in single membership
models. In particular, standard Gibbs sampling approaches tend to get
stuck in local suboptimal configurations where single assignment
changes are not adequate for the identification of probable
alternative configurations \cite{Meeds2007}. Consequently, there is
both a need for computationally efficient models that scale linearly
in the number of links as well as reliable inference schemes for
modeling multiple memberships.

In this paper, we propose a new non-parametric Bayesian latent feature
graph model, denoted the infinite multiple relational model
(\textsc{imrm}), that addresses the challenges mentioned
above. Specifically, the contributions in this paper are the
following: i) We propose the \textsc{imrm} in which inference scales
linearly in the number of links.  ii) We propose a non-conjugate
split-merge sampling procedure for parameter inference.  iii) We demonstrate how the single
membership \textsc{irm} model implicitly accounts for multiple
memberships.  iv) We compare existing non-parametric single membership
models with our proposed multiple membership counterparts in learning
latent structure of a variety of benchmark "real" size networks and demonstrate that explicitly modeling multiple-membership results in more compact representations of latent structure.

\section{Infinite multiple-membership relational model}
Given a graph, assume that each vertex $v_i$ has an associated
$K$-dimensional binary latent feature vector, $\m{z}_{i}$, with
$K_i=|\m{z}_i|_1$ assignments. Consider vertex $v_i$ and $v_j$: For
all $K_iK_j$ combinations of classes there is an associated
probability, $\rho_{k\ell}$, of generating a link. We assume that each
of these combinations of classes act independently to generate a link
between $v_i$ and $v_j$, such that the total probability, $\pi_{ij}$,
of generating a link between $v_i$ and $v_j$ is given by
\begin{equation}
  \label{eq:NoisyOrPi}
  \pi_{ij} = 1-(1-\sigma_{ij})\prod_{k\ell}(1-\rho_{k\ell})^{z_{ik}z_{j\ell}},
\end{equation}
where $\sigma_{ij}$ is an optional term that can be used to account
for noise or to include further side-information as discussed
previously. Under this model, the features act as independent causes
of links, and thus if a vertex gets an additional feature it will
result in an increased probability of linking to other vertices. In
contrast to the model proposed by \cite{Miller2009}, where negative
weights leads to features that inhibit links, our model is more
restricted. Although this might result in less power to explain data,
we expect that it will be easier to interpret the features in our
model because links are directly generated by individual features and
not through complex interactions between features. This is analogous
to non-negative matrix factorization that is known to form parts-based
representation because it does not allow component cancellations
\cite{Lee1999b}. If the latent features $\m{z}_i$ have only a single
active element and $\sigma_{ij}=0$, Eq.~(\ref{eq:NoisyOrPi}) reduces
to $\pi_{ij}=\rho_{c_ic_j}$, i.e., the proposed model directly
generalizes the \textsc{irm} model; hence, we denote our model the
infinite multiple-membership relational model (\textsc{imrm}).

The link probability model in Eq.~(\ref{eq:NoisyOrPi}) has a very
attractive computational property. In many real data sets, the number
of non-links far exceeds the number of links present in the
network. To analyze large scale networks where this holds it is a
great advantage to devise algorithms that scale computationally only
with the number of links present. As we show in the following, our
model has that property. Assuming $\sigma_{ij}=0$ for simplicity of
presentation, we may write Eq.~(\ref{eq:NoisyOrPi}) more compactly as
\newcommand{\zPz}{\m{z}_i^\top\!\!\m{P}\m{z}_j}
\newcommand{\ezPz}{\mathrm{e}^{\zPz}}
$
  \pi_{ij} = 1-\ezPz$,

where the elements of the matrix $\m{P}$ are
$p_{k\ell}=\log(1-\rho_{kl})$. Inserting this in Eq.~(\ref{eq:py}) we have
\begin{equation}
  \label{eq:LikelihoodMIRM}
  p(\m{Y}|\m{Z},\m{P})
  =
  \cprod{(i,j)\in\mathcal{Y}}
  \left(1-\ezPz\right)^{y_{ij}}
  \left(\ezPz\right)^{1-y_{ij}}
  =
  \Big[\!\cprod{(i,j)\in\mathcal{Y}_1}(1-\ezPz)\!\Big]
  \exp\!\Big[\csum{(i,j)\in\mathcal{Y}_0}\zPz\Big]\makebox[0pt]{.}
\end{equation}
The exponent of the second term,
which entails a sum over the possibly large set of non-links in the
network, can be efficiently computed as
\begin{equation}
  \label{eq:SumOverNonLinks}
  \sum_{k\ell}
  p_{k\ell}
  \Big(\sum_{i=1}^Nz_{ik}\sum_{j=1}^Nz_{j\ell}-
    \csum{(i,j)\in\mathcal{Y}_1\cup\mathcal{Y}_?}
    z_{ik}z_{j\ell}\Big),
\end{equation}
requiring only summation over links and ``missing'' links. Assuming
that the graph is not dominated by ``missing'' links, the computation
of Eq.~(\ref{eq:LikelihoodMIRM}) scales linearly in the number of
graph links, $|\mathcal{Y}_1|$. We presently consider latent binary features $\m{z}_{i}$, but we note that the model scales linearly for any parameterizations of the latent feature vector $\m{z}_{i}$, as long as $\pi_{ij}=1-\ezPz \in [0;1]$  which holds in general if $\m{z}_{i}$ is non-negative.

As in existing multiple membership models \cite{Meeds2007,Miller2009}
we will assume an unbounded number of latent features. We learn the
effective number of features through the Indian buffet process
(\textsc{ibp}) representation \cite{Griffiths2006}, which defines a
distribution over unbounded binary matrices,
\begin{equation}
  \label{eq:pz}
  \m{Z} \sim \mathrm{IBP}(\alpha) \propto
  \frac{\alpha^{K}}{\scprod{\m{h}\in[0,1]^N}K_{\m{h}}!}
  \!\prod_{k=1}^{K}\frac{(N-m_k)!(m_k-1)!}{N!}
\end{equation}
where $m_k$ is the number of vertices belonging to class $k$ and
$K_{\m{h}}$ is the number of columns of $\m{Z}$ equal to $\m{h}$.

As a prior over the class link probabilities we choose independent
Beta distributions,
\begin{equation}
  \label{eq:rho}
  \rho_{k\ell}|a_{k\ell},b_{k\ell}\!\sim\! \mathrm{Beta}(a_{k\ell},b_{k\ell})\!\propto\!
  \rho_{k\ell}^{a_{k\ell}-1}(1\!-\!\rho_{k\ell})^{b_{k\ell}-1}\makebox[0pt][c]{.}
\end{equation}
This is a conjugate prior for the single membership models where the
parameters $a_{k\ell}$ and $b_{k\ell}$ correspond to pseudo counts of
links and non-links respectively between classes $k$ and $\ell$.

\subsection{Inference}
\label{sec:Inference}
In the following we present a method for inferring the parameters of
the model: the infinite binary feature matrix $\m{Z}$ and the link
probabilities $\rho_{k\ell}$. In the latent class model when only a
single feature is active for each vertex, the likelihood in
Eq.~(\ref{eq:LikelihoodMIRM}) is conjugate to the Beta prior for
$\rho_{k\ell}$. In that case, $\m{P}$ can be integrated away and a
collapsed Gibbs sampling procedure for $\m{Z}$ can be used
\cite{Kemp2006}. This is not possible in the \textsc{imrm}; instead,
we propose to sample $\m{P}\sim p(\m{P}|\m{Z},\m{Y})$ using
Hamiltonian Markov chain Monte Carlo (\textsc{hmc}), and $\m{Z}\sim
p(\m{Z}|\m{P},\m{Y})$ using Gibbs sampling combined with split-merge
moves.

\textbf{HMC for class link probabilities:} Hamiltonian Markov chain Monte Carlo (\textsc{hmc}) \cite{Duane1987}
is an auxiliary variable sampling procedure that utilizes the gradient
of the log posterior to avoid the random walk behavior of other
sampling methods such as Metropolis-Hastings. In the following we do
not describe the details of the \textsc{hmc} algorithm, but only
derive the required expressions for the gradient. To utilize
\textsc{hmc}, the sampled variables must be unconstrained, but since
$\rho_{k\ell}$ is a probability we make the following change of
variable from $\rho_{k\ell}\in[0,1]$ to
$r_{k\ell}\in(-\infty,\infty)$, $  \rho_{k\ell}=\tfrac{1}{1+\exp(-r_{k\ell})}$, $r_{k\ell}=-\log\left(\rho^{-1}_{k\ell}-1\right)$.
Using the change of variables theorem, the prior for the class link
probabilities expressed in terms of $r_{k\ell}$ is given by $ p(r_{k\ell}|a_{k\ell},b_{k\ell})\propto
  \mathrm{e}^{a_{k\ell}r_{k\ell}}(\mathrm{e}^{r_{k\ell}}+1)^{-(a_{k\ell}+b_{k\ell})}$. With this, the relevant terms of the negative log posterior is given by
\begin{equation}
  \label{eq:LogPosteriorRho}
  -\mathcal{L}_{\m{P}}=\log p(\m{P}|\m{Z},\m{Y}) = c+
  \csum{(i,j)\in\mathcal{Y}_1}\log\left(1-\ezPz\right)+
  \csum{(i,j)\in\mathcal{Y}_0}\zPz
  \!+\!\!\sum_{k\ell}a_{k\ell}r_{k\ell}+(a_{k\ell}\!+\!b_{k\ell})\log(\mathrm{e}^{r_{k\ell}}\!\!+\!1),
\end{equation}
where $c$ does not depend on $\m{P}$. From this, the required gradient can be computed,
\begin{equation}
  \label{eq:GradientRho}
  \frac{\partial\mathcal{L}_{\m{P}}}{\partial r_{k\ell}}=
  -\sum_{(i,j)\in\mathcal{Y}_1}\frac{\ezPz}{1-\ezPz}z_{ik}z_{j\ell}
  \rho_{k\ell}
  +\sum_{(i,j)\in\mathcal{Y}_0}\hspace{-8pt}z_{ik}z_{j\ell} \rho_{k\ell}
  +(a_{k\ell}+b_{k\ell})\rho_{k\ell}-a_{k\ell}.
\end{equation}
Again, the possibly large sum over non-links in the second term can be
computed efficiently as in Eq.~(\ref{eq:SumOverNonLinks}).

\textbf{Gibbs sampler for binary features: }
\label{sec:GibbsSamplerForBinaryFeatures}
Following \cite{Griffiths2006}, a Gibbs sampler for the
latent binary features $\m{Z}$ can be derived. Consider sampling the
$k$th feature of vertex $v_i$: If one or more other vertices also
possess the feature, i.e., $m_{-ik}=\sum_{j\neq i}z_{jk}>0$, the
posterior marginal is given by
\begin{equation}
  \label{eq:GibbsSampler}
  p(z_{ik}=1|\m{Z}_{-(ik)},\m{P},\m{Y})
  \propto p(\m{Y}|\m{Z},\m{P})\frac{m_{-ik}}{N}.
\end{equation}
When evaluating the likelihood term, only the terms that depend on
$z_{ik}$ need be computed and the Gibbs sampler can be implemented
efficiently by reusing computation and by up and down dating
variables.

In addition to sampling existing features,
$K_1^{(i)}=\mathrm{Poisson}\big(\!\tfrac{\alpha}{N}\!\big)$ new
features should also be associated with $v_i$. \cite{Griffiths2006}
suggest {\em``\dots computing probabilities for a range of values of
  $K_1^{(i)}$ up to some reasonable upper bound\dots''}; however,
following \cite{Meeds2007} we take another approach and sample the
new features by Metropolis-Hastings using the prior as proposal
density. The values of $\rho_{kl}$ corresponding to the new features
are proposed from the prior in Eq. (\ref{eq:rho}).

\textbf{Split-merge move for binary features: }
\label{sec:SplitMerge}
A drawback of Gibbs sampling procedures is that only a single variable
is updated at a time, which makes the sampler prone to get stuck in
suboptimal configurations. As a remedy, bolder Metropolis Hasting
moves can be considered in which multiple changes of assignments help
exploring alternative high probability configurations. How these
alternative configurations are proposed is crucial in order to attain
reasonable acceptance rates. A popular approach is to split or merge
existing classes as proposed in \cite{Jain2004} for the Dirichlet
process mixture model (\textsc{dpmm}). Split-merge sampling in the
\textsc{ibp} has previously been discussed briefly by
\cite{Meeds2007} and \cite{Miller2009}.

Inspired by the
non-conjugate sequential allocation split-merge sampler for the
\textsc{dpmm} \cite{Dahl2005}, we propose the following procedure: Draw two non-zero
elements of $\m{Z}$, $(k_1,i_1)$ and $(k_2,i_2)$. If $k_1=k_2$ propose
a split --- otherwise propose to merge classes $k_1$ and $k_2$ into a
joint cluster $k_1$. Accept the proposal with the Metropolis-Hastings
acceptance rate, $
  a^* = \min\left(1,\frac{p(\m{Z}^*,\m{P}^*|\m{Y})q(\m{Z}|\m{Z}^*)q(\m{P}|\m{P}^*)}%
    {p(\m{Z},\m{P}|\m{Y})q(\m{Z}^*|\m{Z})q(\m{P}^*|\m{P})}\right)$.
In case of a merge, we remove $k_2$ and assign all its vertices to
$k_1$, and we remove the corresponding row and column of $\m{P}$ (this
proposal is deterministic and has probability one). For a split, we
remove all vertices except $i_1$ from cluster $k_1=k_2=k$ and create a
new cluster $k^*$ and assign $i_2$ to it. We then sample a new row and
column $\rho^*_{k'\ell'}$ for the new cluster as described below. Next
we sequentially allocate \cite{Dahl2005} the remaining original
members of $k$ to either $k$ or $k^*$ or both in a restricted Gibbs
sampling sweep, and refine the allocation through $t$ additional
restricted Gibbs scans \cite{Jain2004}.

The proposal density for $\rho^*_{k'\ell'}$ is based on a random walk,
$\rho^*_{k'\ell'}\sim\mathrm{Beta}(\bar a_{k'\ell'},\bar
b_{k'\ell'})$, where
\begin{equation}
  \label{eq:RhoProposal}
  \bar b_{k'\ell'}=\max\big(1,(1-\bar\rho_{k'\ell'})m_{k}^2-1+\bar\rho_{k'\ell'}\big),\quad
  \bar a_{k'\ell'}=\max\Big(1,\frac{\bar\rho_{k'\ell'}}{1-\bar\rho_{k'\ell'}}\bar b_{k'\ell'}\Big),
\end{equation}
such that $\rho^*_{k'\ell'}$ has mean $\bar\rho_{k'\ell'}$ and
variance equal to the empirical variance,
$\bar\rho_{k'\ell'}(1-\bar\rho_{k'\ell'})/m_{k}^2$. We choose the mean
of the random walk as $
\bar\rho_{k'\ell'}=\left\{
    \begin{array}{ll}
      \rho_{kk} & \ell'=k'=k^*\\
      \tfrac{1}{K-1}\sum_{\ell\neq k}\rho_{k\ell} & k'=k^*, \ell'=k\\
      \tfrac{1}{K-1}\sum_{\ell\neq k}\rho_{\ell k} & k'=k, \ell'=k^*\\
      \rho_{k\ell} & \mathrm{otherwise},\\
    \end{array}
  \right.
$ such that the new class has a similar within and between class link
probabilities as the original class, but such that the class link
probability between the original and new cluster is similar to the
remaining between class link probabilities. This choice is crucial,
since it favors splitting classes into two classes that are no more
related than the relation to the remaining classes.

\section{Results}
Based on the \textsc{hw}, \textsc{db}, and \textsc{rm} parametrization
of $\m{\rho}$, we compared our proposed \textsc{imrm} to the
corresponding single-membership \textsc{irm} \cite{Kemp2006}. We
evaluated the models on a range of synthetically generated as well as
real world networks. We assessed model performance in terms of ability
to predict held-out links and non-links. As performance measure we
used the area under curve (\textsc{auc}) of the receiver operating
characteristic (\textsc{roc}). We also computed the predictive log
likelihood (not shown here) which gave similar results. For
comparison, we included the performance of several standard
non-parametric link prediction approaches based on the following
scores,
\begin{equation*} \gamma_{i,j}^{\mathrm{ComN}}=\boldsymbol{y}_i^\top
  \boldsymbol{y}_j,\quad
  \gamma_{i,j}^{\mathrm{DegPr}}=k_ik_j,\quad
  \gamma_{i,j}^{\mathrm{Jacc}}=\frac{\boldsymbol{y}_i^\top
    \boldsymbol{y}_j}{k_i+ k_j-\boldsymbol{y}_i^\top
    \boldsymbol{y}_j},\quad
  \gamma_{i,j}^{\mathrm{ShP}}=\frac{1}{\min_p\{(\boldsymbol{Y}^p)_{i,j}>0\}},
\end{equation*}
where $k_i=\sum_{j} y_{ij}$ is the degree of vertex $v_i$.

In all the analyses we removed $2.5\%$ of the links and an equivalent
number of non-links for cross-validation. We analyzed a total of five
random data splits and all of the analyses were based on 2500 sampling
iterations initialized randomly with $K=50$ clusters. Each iteration
was based on split-merge sampling using sequential allocation with
$t=2$ restricted Gibbs scans followed by standard Gibbs sampling. Our
implementation of the \textsc{irm} was based on collapsed Gibbs
sampling (i.e. integrating out $\boldsymbol{\rho}$) as proposed in
\cite{Kemp2006} but we also included a conjugate single-membership
split-merge step corresponding to the proposed non-conjugate
split-merge sampler. The priors were chosen as $\alpha=\log(N)$,
$a_{kk}=5$, $a_{k\ell}=1\forall k\neq\ell$ and $b_{kk}=1$,
$b_{k\ell}=5\forall k\neq\ell$ which renders the priors practically
non-informative.

\textbf{Synthetic networks: } We analyzed a total of six synthetic networks generated according to
the \textsc{hw}, \textsc{db} and \textsc{rm} models based on the
vertices having either one or two memberships to the underlying
classes. For the single membership models we generated a total of
$K=5$ groups each containing 100 vertices. For the \textsc{hw}
generated network we set $\rho_{c}=1$ and $\rho_0=0$ while for the
\textsc{db} generated network we used a within community densities
$\rho_{k}$ ranging from 0.2 to 1 while $\rho_0=0$. The \textsc{rm}
generated network had same within community densities as the
\textsc{db} network but included varying degrees of overlap between
the communities. The multiple membership models
denoted \textsc{mhw}, \textsc{mdb} and \textsc{mrm} were
generated from the corresponding single membership models as $\m{Y}\vee
\m{R}\m{Y}\m{R}^\top$ (where $\vee$ denotes element-wise or and
$\m{R}$ is a random permutation matrix with diagonal zero), such that
each vertex belongs to two classes.
\newcommand{\dsc}[1]{\multicolumn{6}{l}{\tiny\rm\vspace{2pt}\hspace{10pt}#1}}
\newcommand{\dci}[1]{\cite{#1}\vspace{0pt}}
\newcommand{\pts}{\textperthousand}

\begin{figure*}[tb]
  \centering
%
%
\begin{psfrags}%
\psfragscanon%
%
\psfrag{s18}[b][b]{\color[rgb]{0,0,0}\setlength{\tabcolsep}{0pt}\begin{tabular}{c}$\m{Y}$\end{tabular}}%
\psfrag{s19}[b][b]{\color[rgb]{0,0,0}\setlength{\tabcolsep}{0pt}\begin{tabular}{c}$\m{Z}$\end{tabular}}%
\psfrag{s20}[b][b]{\color[rgb]{0,0,0}\setlength{\tabcolsep}{0pt}\begin{tabular}{c}$\m{\rho}$\end{tabular}}%
%
\resizebox{5.5cm}{!}{\includegraphics{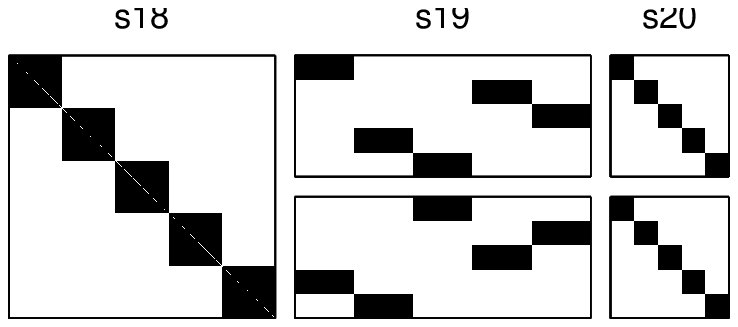}}%
\end{psfrags}%
%
\quad
%
%
\begin{psfrags}%
\psfragscanon%
%
\psfrag{s18}[b][b]{\color[rgb]{0,0,0}\setlength{\tabcolsep}{0pt}\begin{tabular}{c}$\m{Y}$\end{tabular}}%
\psfrag{s19}[b][b]{\color[rgb]{0,0,0}\setlength{\tabcolsep}{0pt}\begin{tabular}{c}$\m{Z}$\end{tabular}}%
\psfrag{s20}[b][b]{\color[rgb]{0,0,0}\setlength{\tabcolsep}{0pt}\begin{tabular}{c}$\m{\rho}$\end{tabular}}%
%
\resizebox{5.5cm}{!}{\includegraphics{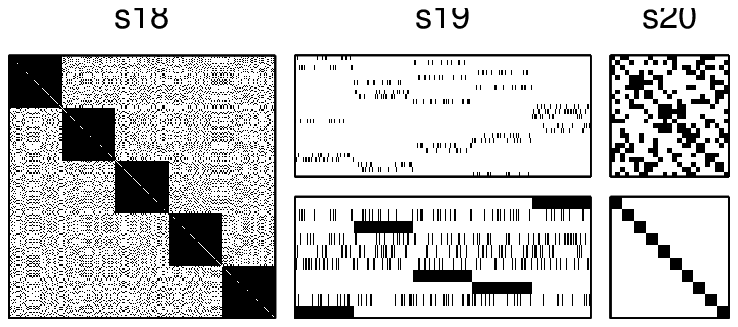}}%
\end{psfrags}%
%

  \caption{\textsc{irm} (upper) and \textsc{imrm} (lower) analysis of
    single (left) and multiple membership \textsc{hw} network
    (right). On the single membership data, both models find the
    correct class assignments. On the multiple membership data, the
    \textsc{imrm} finds the correct 10 classes, while \textsc{irm}
    extracts 25 classes, which through $\m{\rho}$ accounts for all
    combinations of classes present in the data.}
  \label{fig:SynthExample}
\end{figure*}

\begin{figure*}[tb]
  \centering
%
%
\begin{psfrags}%
\psfragscanon%
%
\psfrag{s13}[b][b]{\color[rgb]{0,0,0}\setlength{\tabcolsep}{0pt}\begin{tabular}{c}1\end{tabular}}%
\psfrag{s14}[b][b]{\color[rgb]{0,0,0}\setlength{\tabcolsep}{0pt}\begin{tabular}{c}2\end{tabular}}%
\psfrag{s15}[b][b]{\color[rgb]{0,0,0}\setlength{\tabcolsep}{0pt}\begin{tabular}{c}3\end{tabular}}%
\psfrag{s16}[b][b]{\color[rgb]{0,0,0}\setlength{\tabcolsep}{0pt}\begin{tabular}{c}4\end{tabular}}%
\psfrag{s17}[b][b]{\color[rgb]{0,0,0}\setlength{\tabcolsep}{0pt}\begin{tabular}{c}5\end{tabular}}%
\psfrag{s18}[b][b]{\color[rgb]{0,0,0}\setlength{\tabcolsep}{0pt}\begin{tabular}{c}6\end{tabular}}%
\psfrag{s19}[b][b]{\color[rgb]{0,0,0}\setlength{\tabcolsep}{0pt}\begin{tabular}{c}7\end{tabular}}%
\psfrag{s20}[b][b]{\color[rgb]{0,0,0}\setlength{\tabcolsep}{0pt}\begin{tabular}{c}8\end{tabular}}%
\psfrag{s21}[b][b]{\color[rgb]{0,0,0}\setlength{\tabcolsep}{0pt}\begin{tabular}{c}9\end{tabular}}%
\psfrag{s22}[b][b]{\color[rgb]{0,0,0}\setlength{\tabcolsep}{0pt}\begin{tabular}{c}10\end{tabular}}%
\psfrag{s24}[b][b]{\color[rgb]{0,0,0}\setlength{\tabcolsep}{0pt}\begin{tabular}{c}DB\end{tabular}}%
\psfrag{s25}[b][b]{\color[rgb]{0,0,0}\setlength{\tabcolsep}{0pt}\begin{tabular}{c}1\end{tabular}}%
\psfrag{s26}[b][b]{\color[rgb]{0,0,0}\setlength{\tabcolsep}{0pt}\begin{tabular}{c}2\end{tabular}}%
\psfrag{s27}[b][b]{\color[rgb]{0,0,0}\setlength{\tabcolsep}{0pt}\begin{tabular}{c}3\end{tabular}}%
\psfrag{s28}[b][b]{\color[rgb]{0,0,0}\setlength{\tabcolsep}{0pt}\begin{tabular}{c}4\end{tabular}}%
\psfrag{s29}[b][b]{\color[rgb]{0,0,0}\setlength{\tabcolsep}{0pt}\begin{tabular}{c}5\end{tabular}}%
\psfrag{s30}[b][b]{\color[rgb]{0,0,0}\setlength{\tabcolsep}{0pt}\begin{tabular}{c}6\end{tabular}}%
\psfrag{s31}[b][b]{\color[rgb]{0,0,0}\setlength{\tabcolsep}{0pt}\begin{tabular}{c}7\end{tabular}}%
\psfrag{s32}[b][b]{\color[rgb]{0,0,0}\setlength{\tabcolsep}{0pt}\begin{tabular}{c}8\end{tabular}}%
\psfrag{s33}[b][b]{\color[rgb]{0,0,0}\setlength{\tabcolsep}{0pt}\begin{tabular}{c}9\end{tabular}}%
\psfrag{s34}[b][b]{\color[rgb]{0,0,0}\setlength{\tabcolsep}{0pt}\begin{tabular}{c}10\end{tabular}}%
\psfrag{s36}[b][b]{\color[rgb]{0,0,0}\setlength{\tabcolsep}{0pt}\begin{tabular}{c}RM\end{tabular}}%
\psfrag{s37}[b][b]{\color[rgb]{0,0,0}\setlength{\tabcolsep}{0pt}\begin{tabular}{c}1\end{tabular}}%
\psfrag{s38}[b][b]{\color[rgb]{0,0,0}\setlength{\tabcolsep}{0pt}\begin{tabular}{c}2\end{tabular}}%
\psfrag{s39}[b][b]{\color[rgb]{0,0,0}\setlength{\tabcolsep}{0pt}\begin{tabular}{c}3\end{tabular}}%
\psfrag{s40}[b][b]{\color[rgb]{0,0,0}\setlength{\tabcolsep}{0pt}\begin{tabular}{c}4\end{tabular}}%
\psfrag{s41}[b][b]{\color[rgb]{0,0,0}\setlength{\tabcolsep}{0pt}\begin{tabular}{c}5\end{tabular}}%
\psfrag{s42}[b][b]{\color[rgb]{0,0,0}\setlength{\tabcolsep}{0pt}\begin{tabular}{c}6\end{tabular}}%
\psfrag{s43}[b][b]{\color[rgb]{0,0,0}\setlength{\tabcolsep}{0pt}\begin{tabular}{c}7\end{tabular}}%
\psfrag{s44}[b][b]{\color[rgb]{0,0,0}\setlength{\tabcolsep}{0pt}\begin{tabular}{c}8\end{tabular}}%
\psfrag{s45}[b][b]{\color[rgb]{0,0,0}\setlength{\tabcolsep}{0pt}\begin{tabular}{c}9\end{tabular}}%
\psfrag{s46}[b][b]{\color[rgb]{0,0,0}\setlength{\tabcolsep}{0pt}\begin{tabular}{c}10\end{tabular}}%
\psfrag{s48}[b][b]{\color[rgb]{0,0,0}\setlength{\tabcolsep}{0pt}\begin{tabular}{c}MHW\end{tabular}}%
\psfrag{s49}[b][b]{\color[rgb]{0,0,0}\setlength{\tabcolsep}{0pt}\begin{tabular}{c}1\end{tabular}}%
\psfrag{s50}[b][b]{\color[rgb]{0,0,0}\setlength{\tabcolsep}{0pt}\begin{tabular}{c}2\end{tabular}}%
\psfrag{s51}[b][b]{\color[rgb]{0,0,0}\setlength{\tabcolsep}{0pt}\begin{tabular}{c}3\end{tabular}}%
\psfrag{s52}[b][b]{\color[rgb]{0,0,0}\setlength{\tabcolsep}{0pt}\begin{tabular}{c}4\end{tabular}}%
\psfrag{s53}[b][b]{\color[rgb]{0,0,0}\setlength{\tabcolsep}{0pt}\begin{tabular}{c}5\end{tabular}}%
\psfrag{s54}[b][b]{\color[rgb]{0,0,0}\setlength{\tabcolsep}{0pt}\begin{tabular}{c}6\end{tabular}}%
\psfrag{s55}[b][b]{\color[rgb]{0,0,0}\setlength{\tabcolsep}{0pt}\begin{tabular}{c}7\end{tabular}}%
\psfrag{s56}[b][b]{\color[rgb]{0,0,0}\setlength{\tabcolsep}{0pt}\begin{tabular}{c}8\end{tabular}}%
\psfrag{s57}[b][b]{\color[rgb]{0,0,0}\setlength{\tabcolsep}{0pt}\begin{tabular}{c}9\end{tabular}}%
\psfrag{s58}[b][b]{\color[rgb]{0,0,0}\setlength{\tabcolsep}{0pt}\begin{tabular}{c}10\end{tabular}}%
\psfrag{s60}[b][b]{\color[rgb]{0,0,0}\setlength{\tabcolsep}{0pt}\begin{tabular}{c}MDB\end{tabular}}%
\psfrag{s61}[b][b]{\color[rgb]{0,0,0}\setlength{\tabcolsep}{0pt}\begin{tabular}{c}1\end{tabular}}%
\psfrag{s62}[b][b]{\color[rgb]{0,0,0}\setlength{\tabcolsep}{0pt}\begin{tabular}{c}2\end{tabular}}%
\psfrag{s63}[b][b]{\color[rgb]{0,0,0}\setlength{\tabcolsep}{0pt}\begin{tabular}{c}3\end{tabular}}%
\psfrag{s64}[b][b]{\color[rgb]{0,0,0}\setlength{\tabcolsep}{0pt}\begin{tabular}{c}4\end{tabular}}%
\psfrag{s65}[b][b]{\color[rgb]{0,0,0}\setlength{\tabcolsep}{0pt}\begin{tabular}{c}5\end{tabular}}%
\psfrag{s66}[b][b]{\color[rgb]{0,0,0}\setlength{\tabcolsep}{0pt}\begin{tabular}{c}6\end{tabular}}%
\psfrag{s67}[b][b]{\color[rgb]{0,0,0}\setlength{\tabcolsep}{0pt}\begin{tabular}{c}7\end{tabular}}%
\psfrag{s68}[b][b]{\color[rgb]{0,0,0}\setlength{\tabcolsep}{0pt}\begin{tabular}{c}8\end{tabular}}%
\psfrag{s69}[b][b]{\color[rgb]{0,0,0}\setlength{\tabcolsep}{0pt}\begin{tabular}{c}9\end{tabular}}%
\psfrag{s70}[b][b]{\color[rgb]{0,0,0}\setlength{\tabcolsep}{0pt}\begin{tabular}{c}10\end{tabular}}%
\psfrag{s72}[b][b]{\color[rgb]{0,0,0}\setlength{\tabcolsep}{0pt}\begin{tabular}{c}MRM\end{tabular}}%
\psfrag{s73}[b][b]{\color[rgb]{0,0,0}\setlength{\tabcolsep}{0pt}\begin{tabular}{c}1\end{tabular}}%
\psfrag{s74}[b][b]{\color[rgb]{0,0,0}\setlength{\tabcolsep}{0pt}\begin{tabular}{c}2\end{tabular}}%
\psfrag{s75}[b][b]{\color[rgb]{0,0,0}\setlength{\tabcolsep}{0pt}\begin{tabular}{c}3\end{tabular}}%
\psfrag{s76}[b][b]{\color[rgb]{0,0,0}\setlength{\tabcolsep}{0pt}\begin{tabular}{c}4\end{tabular}}%
\psfrag{s77}[b][b]{\color[rgb]{0,0,0}\setlength{\tabcolsep}{0pt}\begin{tabular}{c}5\end{tabular}}%
\psfrag{s78}[b][b]{\color[rgb]{0,0,0}\setlength{\tabcolsep}{0pt}\begin{tabular}{c}6\end{tabular}}%
\psfrag{s79}[b][b]{\color[rgb]{0,0,0}\setlength{\tabcolsep}{0pt}\begin{tabular}{c}7\end{tabular}}%
\psfrag{s80}[b][b]{\color[rgb]{0,0,0}\setlength{\tabcolsep}{0pt}\begin{tabular}{c}8\end{tabular}}%
\psfrag{s81}[b][b]{\color[rgb]{0,0,0}\setlength{\tabcolsep}{0pt}\begin{tabular}{c}9\end{tabular}}%
\psfrag{s82}[b][b]{\color[rgb]{0,0,0}\setlength{\tabcolsep}{0pt}\begin{tabular}{c}10\end{tabular}}%
\psfrag{s84}[b][b]{\color[rgb]{0,0,0}\setlength{\tabcolsep}{0pt}\begin{tabular}{c}HW\end{tabular}}%
%
\psfrag{x01}[t][t]{\rotatebox{90}{\tiny\sc ComN}}%
\psfrag{x02}[t][t]{\rotatebox{90}{\tiny\sc Jac}}%
\psfrag{x03}[t][t]{\rotatebox{90}{\tiny\sc DegP}}%
\psfrag{x04}[t][t]{\rotatebox{90}{\tiny\sc ShP}}%
\psfrag{x05}[t][t]{\rotatebox{90}{\tiny\sc IHW}}%
\psfrag{x06}[t][t]{\rotatebox{90}{\tiny\sc IDB}}%
\psfrag{x07}[t][t]{\rotatebox{90}{\tiny\sc IRM}}%
\psfrag{x08}[t][t]{\rotatebox{90}{\tiny\sc IMHW}}%
\psfrag{x09}[t][t]{\rotatebox{90}{\tiny\sc IMDB}}%
\psfrag{x10}[t][t]{\rotatebox{90}{\tiny\sc IMRM}}%
\psfrag{x11}[t][t]{\rotatebox{90}{\tiny\sc ComN}}%
\psfrag{x12}[t][t]{\rotatebox{90}{\tiny\sc Jac}}%
\psfrag{x13}[t][t]{\rotatebox{90}{\tiny\sc DegP}}%
\psfrag{x14}[t][t]{\rotatebox{90}{\tiny\sc ShP}}%
\psfrag{x15}[t][t]{\rotatebox{90}{\tiny\sc IHW}}%
\psfrag{x16}[t][t]{\rotatebox{90}{\tiny\sc IDB}}%
\psfrag{x17}[t][t]{\rotatebox{90}{\tiny\sc IRM}}%
\psfrag{x18}[t][t]{\rotatebox{90}{\tiny\sc IMHW}}%
\psfrag{x19}[t][t]{\rotatebox{90}{\tiny\sc IMDB}}%
\psfrag{x20}[t][t]{\rotatebox{90}{\tiny\sc IMRM}}%
\psfrag{x21}[t][t]{\rotatebox{90}{\tiny\sc ComN}}%
\psfrag{x22}[t][t]{\rotatebox{90}{\tiny\sc Jac}}%
\psfrag{x23}[t][t]{\rotatebox{90}{\tiny\sc DegP}}%
\psfrag{x24}[t][t]{\rotatebox{90}{\tiny\sc ShP}}%
\psfrag{x25}[t][t]{\rotatebox{90}{\tiny\sc IHW}}%
\psfrag{x26}[t][t]{\rotatebox{90}{\tiny\sc IDB}}%
\psfrag{x27}[t][t]{\rotatebox{90}{\tiny\sc IRM}}%
\psfrag{x28}[t][t]{\rotatebox{90}{\tiny\sc IMHW}}%
\psfrag{x29}[t][t]{\rotatebox{90}{\tiny\sc IMDB}}%
\psfrag{x30}[t][t]{\rotatebox{90}{\tiny\sc IMRM}}%
\psfrag{x31}[t][t]{\rotatebox{90}{\tiny\sc ComN}}%
\psfrag{x32}[t][t]{\rotatebox{90}{\tiny\sc Jac}}%
\psfrag{x33}[t][t]{\rotatebox{90}{\tiny\sc DegP}}%
\psfrag{x34}[t][t]{\rotatebox{90}{\tiny\sc ShP}}%
\psfrag{x35}[t][t]{\rotatebox{90}{\tiny\sc IHW}}%
\psfrag{x36}[t][t]{\rotatebox{90}{\tiny\sc IDB}}%
\psfrag{x37}[t][t]{\rotatebox{90}{\tiny\sc IRM}}%
\psfrag{x38}[t][t]{\rotatebox{90}{\tiny\sc IMHW}}%
\psfrag{x39}[t][t]{\rotatebox{90}{\tiny\sc IMDB}}%
\psfrag{x40}[t][t]{\rotatebox{90}{\tiny\sc IMRM}}%
\psfrag{x41}[t][t]{\rotatebox{90}{\tiny\sc ComN}}%
\psfrag{x42}[t][t]{\rotatebox{90}{\tiny\sc Jac}}%
\psfrag{x43}[t][t]{\rotatebox{90}{\tiny\sc DegP}}%
\psfrag{x44}[t][t]{\rotatebox{90}{\tiny\sc ShP}}%
\psfrag{x45}[t][t]{\rotatebox{90}{\tiny\sc IHW}}%
\psfrag{x46}[t][t]{\rotatebox{90}{\tiny\sc IDB}}%
\psfrag{x47}[t][t]{\rotatebox{90}{\tiny\sc IRM}}%
\psfrag{x48}[t][t]{\rotatebox{90}{\tiny\sc IMHW}}%
\psfrag{x49}[t][t]{\rotatebox{90}{\tiny\sc IMDB}}%
\psfrag{x50}[t][t]{\rotatebox{90}{\tiny\sc IMRM}}%
\psfrag{x51}[t][t]{\rotatebox{90}{\tiny\sc ComN}}%
\psfrag{x52}[t][t]{\rotatebox{90}{\tiny\sc Jac}}%
\psfrag{x53}[t][t]{\rotatebox{90}{\tiny\sc DegP}}%
\psfrag{x54}[t][t]{\rotatebox{90}{\tiny\sc ShP}}%
\psfrag{x55}[t][t]{\rotatebox{90}{\tiny\sc IHW}}%
\psfrag{x56}[t][t]{\rotatebox{90}{\tiny\sc IDB}}%
\psfrag{x57}[t][t]{\rotatebox{90}{\tiny\sc IRM}}%
\psfrag{x58}[t][t]{\rotatebox{90}{\tiny\sc IMHW}}%
\psfrag{x59}[t][t]{\rotatebox{90}{\tiny\sc IMDB}}%
\psfrag{x60}[t][t]{\rotatebox{90}{\tiny\sc IMRM}}%
%
\psfrag{v01}[r][r]{0.4}%
\psfrag{v02}[r][r]{0.6}%
\psfrag{v03}[r][r]{0.8}%
\psfrag{v04}[r][r]{1}%
%
\resizebox{12cm}{!}{\includegraphics{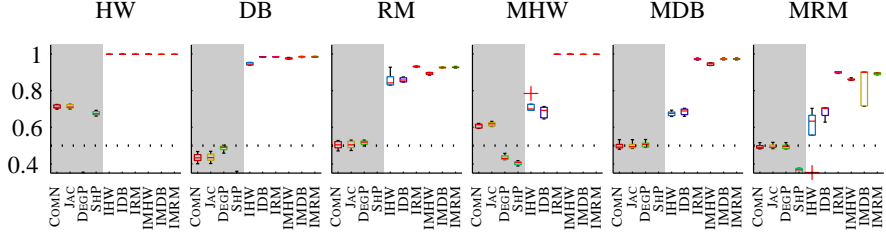}}%
\end{psfrags}%
%
\\
  \vspace{5pt}
  \caption{\textsc{auc} scores for the analysis of the six synthetically generated data sets.}
  \label{fig:SynthResults}
\end{figure*}

\begin{figure*}[tbp]
  \centering
%
%
\begin{psfrags}%
\psfragscanon%
%
\psfrag{s11}[b][b]{\color[rgb]{0,0,0}\setlength{\tabcolsep}{0pt}\begin{tabular}{c}1\end{tabular}}%
\psfrag{s12}[b][b]{\color[rgb]{0,0,0}\setlength{\tabcolsep}{0pt}\begin{tabular}{c}2\end{tabular}}%
\psfrag{s13}[b][b]{\color[rgb]{0,0,0}\setlength{\tabcolsep}{0pt}\begin{tabular}{c}3\end{tabular}}%
\psfrag{s14}[b][b]{\color[rgb]{0,0,0}\setlength{\tabcolsep}{0pt}\begin{tabular}{c}4\end{tabular}}%
\psfrag{s15}[b][b]{\color[rgb]{0,0,0}\setlength{\tabcolsep}{0pt}\begin{tabular}{c}5\end{tabular}}%
\psfrag{s16}[b][b]{\color[rgb]{0,0,0}\setlength{\tabcolsep}{0pt}\begin{tabular}{c}6\end{tabular}}%
\psfrag{s17}[b][b]{\color[rgb]{0,0,0}\setlength{\tabcolsep}{0pt}\begin{tabular}{c}7\end{tabular}}%
\psfrag{s18}[b][b]{\color[rgb]{0,0,0}\setlength{\tabcolsep}{0pt}\begin{tabular}{c}8\end{tabular}}%
\psfrag{s19}[b][b]{\color[rgb]{0,0,0}\setlength{\tabcolsep}{0pt}\begin{tabular}{c}9\end{tabular}}%
\psfrag{s20}[b][b]{\color[rgb]{0,0,0}\setlength{\tabcolsep}{0pt}\begin{tabular}{c}10\end{tabular}}%
\psfrag{s22}[b][b]{\color[rgb]{0,0,0}\setlength{\tabcolsep}{0pt}\begin{tabular}{c}USPower\end{tabular}}%
\psfrag{s23}[b][b]{\color[rgb]{0,0,0}\setlength{\tabcolsep}{0pt}\begin{tabular}{c}1\end{tabular}}%
\psfrag{s24}[b][b]{\color[rgb]{0,0,0}\setlength{\tabcolsep}{0pt}\begin{tabular}{c}2\end{tabular}}%
\psfrag{s25}[b][b]{\color[rgb]{0,0,0}\setlength{\tabcolsep}{0pt}\begin{tabular}{c}3\end{tabular}}%
\psfrag{s26}[b][b]{\color[rgb]{0,0,0}\setlength{\tabcolsep}{0pt}\begin{tabular}{c}4\end{tabular}}%
\psfrag{s27}[b][b]{\color[rgb]{0,0,0}\setlength{\tabcolsep}{0pt}\begin{tabular}{c}5\end{tabular}}%
\psfrag{s28}[b][b]{\color[rgb]{0,0,0}\setlength{\tabcolsep}{0pt}\begin{tabular}{c}6\end{tabular}}%
\psfrag{s29}[b][b]{\color[rgb]{0,0,0}\setlength{\tabcolsep}{0pt}\begin{tabular}{c}7\end{tabular}}%
\psfrag{s30}[b][b]{\color[rgb]{0,0,0}\setlength{\tabcolsep}{0pt}\begin{tabular}{c}8\end{tabular}}%
\psfrag{s31}[b][b]{\color[rgb]{0,0,0}\setlength{\tabcolsep}{0pt}\begin{tabular}{c}9\end{tabular}}%
\psfrag{s32}[b][b]{\color[rgb]{0,0,0}\setlength{\tabcolsep}{0pt}\begin{tabular}{c}10\end{tabular}}%
\psfrag{s34}[b][b]{\color[rgb]{0,0,0}\setlength{\tabcolsep}{0pt}\begin{tabular}{c}Erdos\end{tabular}}%
\psfrag{s35}[b][b]{\color[rgb]{0,0,0}\setlength{\tabcolsep}{0pt}\begin{tabular}{c}1\end{tabular}}%
\psfrag{s36}[b][b]{\color[rgb]{0,0,0}\setlength{\tabcolsep}{0pt}\begin{tabular}{c}2\end{tabular}}%
\psfrag{s37}[b][b]{\color[rgb]{0,0,0}\setlength{\tabcolsep}{0pt}\begin{tabular}{c}3\end{tabular}}%
\psfrag{s38}[b][b]{\color[rgb]{0,0,0}\setlength{\tabcolsep}{0pt}\begin{tabular}{c}4\end{tabular}}%
\psfrag{s39}[b][b]{\color[rgb]{0,0,0}\setlength{\tabcolsep}{0pt}\begin{tabular}{c}5\end{tabular}}%
\psfrag{s40}[b][b]{\color[rgb]{0,0,0}\setlength{\tabcolsep}{0pt}\begin{tabular}{c}6\end{tabular}}%
\psfrag{s41}[b][b]{\color[rgb]{0,0,0}\setlength{\tabcolsep}{0pt}\begin{tabular}{c}7\end{tabular}}%
\psfrag{s42}[b][b]{\color[rgb]{0,0,0}\setlength{\tabcolsep}{0pt}\begin{tabular}{c}8\end{tabular}}%
\psfrag{s43}[b][b]{\color[rgb]{0,0,0}\setlength{\tabcolsep}{0pt}\begin{tabular}{c}9\end{tabular}}%
\psfrag{s44}[b][b]{\color[rgb]{0,0,0}\setlength{\tabcolsep}{0pt}\begin{tabular}{c}10\end{tabular}}%
\psfrag{s46}[b][b]{\color[rgb]{0,0,0}\setlength{\tabcolsep}{0pt}\begin{tabular}{c}FreeAssoc\end{tabular}}%
\psfrag{s47}[b][b]{\color[rgb]{0,0,0}\setlength{\tabcolsep}{0pt}\begin{tabular}{c}1\end{tabular}}%
\psfrag{s48}[b][b]{\color[rgb]{0,0,0}\setlength{\tabcolsep}{0pt}\begin{tabular}{c}2\end{tabular}}%
\psfrag{s49}[b][b]{\color[rgb]{0,0,0}\setlength{\tabcolsep}{0pt}\begin{tabular}{c}3\end{tabular}}%
\psfrag{s50}[b][b]{\color[rgb]{0,0,0}\setlength{\tabcolsep}{0pt}\begin{tabular}{c}4\end{tabular}}%
\psfrag{s51}[b][b]{\color[rgb]{0,0,0}\setlength{\tabcolsep}{0pt}\begin{tabular}{c}5\end{tabular}}%
\psfrag{s52}[b][b]{\color[rgb]{0,0,0}\setlength{\tabcolsep}{0pt}\begin{tabular}{c}6\end{tabular}}%
\psfrag{s53}[b][b]{\color[rgb]{0,0,0}\setlength{\tabcolsep}{0pt}\begin{tabular}{c}7\end{tabular}}%
\psfrag{s54}[b][b]{\color[rgb]{0,0,0}\setlength{\tabcolsep}{0pt}\begin{tabular}{c}8\end{tabular}}%
\psfrag{s55}[b][b]{\color[rgb]{0,0,0}\setlength{\tabcolsep}{0pt}\begin{tabular}{c}9\end{tabular}}%
\psfrag{s56}[b][b]{\color[rgb]{0,0,0}\setlength{\tabcolsep}{0pt}\begin{tabular}{c}10\end{tabular}}%
\psfrag{s58}[b][b]{\color[rgb]{0,0,0}\setlength{\tabcolsep}{0pt}\begin{tabular}{c}Reuters911\end{tabular}}%
\psfrag{s59}[b][b]{\color[rgb]{0,0,0}\setlength{\tabcolsep}{0pt}\begin{tabular}{c}1\end{tabular}}%
\psfrag{s60}[b][b]{\color[rgb]{0,0,0}\setlength{\tabcolsep}{0pt}\begin{tabular}{c}2\end{tabular}}%
\psfrag{s61}[b][b]{\color[rgb]{0,0,0}\setlength{\tabcolsep}{0pt}\begin{tabular}{c}3\end{tabular}}%
\psfrag{s62}[b][b]{\color[rgb]{0,0,0}\setlength{\tabcolsep}{0pt}\begin{tabular}{c}4\end{tabular}}%
\psfrag{s63}[b][b]{\color[rgb]{0,0,0}\setlength{\tabcolsep}{0pt}\begin{tabular}{c}5\end{tabular}}%
\psfrag{s64}[b][b]{\color[rgb]{0,0,0}\setlength{\tabcolsep}{0pt}\begin{tabular}{c}6\end{tabular}}%
\psfrag{s65}[b][b]{\color[rgb]{0,0,0}\setlength{\tabcolsep}{0pt}\begin{tabular}{c}7\end{tabular}}%
\psfrag{s66}[b][b]{\color[rgb]{0,0,0}\setlength{\tabcolsep}{0pt}\begin{tabular}{c}8\end{tabular}}%
\psfrag{s67}[b][b]{\color[rgb]{0,0,0}\setlength{\tabcolsep}{0pt}\begin{tabular}{c}9\end{tabular}}%
\psfrag{s68}[b][b]{\color[rgb]{0,0,0}\setlength{\tabcolsep}{0pt}\begin{tabular}{c}10\end{tabular}}%
\psfrag{s70}[b][b]{\color[rgb]{0,0,0}\setlength{\tabcolsep}{0pt}\begin{tabular}{c}Yeast\end{tabular}}%
%
\psfrag{x01}[t][t]{\rotatebox{90}{\tiny\sc ComN}}%
\psfrag{x02}[t][t]{\rotatebox{90}{\tiny\sc Jac}}%
\psfrag{x03}[t][t]{\rotatebox{90}{\tiny\sc DegP}}%
\psfrag{x04}[t][t]{\rotatebox{90}{\tiny\sc ShP}}%
\psfrag{x05}[t][t]{\rotatebox{90}{\tiny\sc IHW}}%
\psfrag{x06}[t][t]{\rotatebox{90}{\tiny\sc IDB}}%
\psfrag{x07}[t][t]{\rotatebox{90}{\tiny\sc IRM}}%
\psfrag{x08}[t][t]{\rotatebox{90}{\tiny\sc IMHW}}%
\psfrag{x09}[t][t]{\rotatebox{90}{\tiny\sc IMDB}}%
\psfrag{x10}[t][t]{\rotatebox{90}{\tiny\sc IMRM}}%
\psfrag{x11}[t][t]{\rotatebox{90}{\tiny\sc ComN}}%
\psfrag{x12}[t][t]{\rotatebox{90}{\tiny\sc Jac}}%
\psfrag{x13}[t][t]{\rotatebox{90}{\tiny\sc DegP}}%
\psfrag{x14}[t][t]{\rotatebox{90}{\tiny\sc ShP}}%
\psfrag{x15}[t][t]{\rotatebox{90}{\tiny\sc IHW}}%
\psfrag{x16}[t][t]{\rotatebox{90}{\tiny\sc IDB}}%
\psfrag{x17}[t][t]{\rotatebox{90}{\tiny\sc IRM}}%
\psfrag{x18}[t][t]{\rotatebox{90}{\tiny\sc IMHW}}%
\psfrag{x19}[t][t]{\rotatebox{90}{\tiny\sc IMDB}}%
\psfrag{x20}[t][t]{\rotatebox{90}{\tiny\sc IMRM}}%
\psfrag{x21}[t][t]{\rotatebox{90}{\tiny\sc ComN}}%
\psfrag{x22}[t][t]{\rotatebox{90}{\tiny\sc Jac}}%
\psfrag{x23}[t][t]{\rotatebox{90}{\tiny\sc DegP}}%
\psfrag{x24}[t][t]{\rotatebox{90}{\tiny\sc ShP}}%
\psfrag{x25}[t][t]{\rotatebox{90}{\tiny\sc IHW}}%
\psfrag{x26}[t][t]{\rotatebox{90}{\tiny\sc IDB}}%
\psfrag{x27}[t][t]{\rotatebox{90}{\tiny\sc IRM}}%
\psfrag{x28}[t][t]{\rotatebox{90}{\tiny\sc IMHW}}%
\psfrag{x29}[t][t]{\rotatebox{90}{\tiny\sc IMDB}}%
\psfrag{x30}[t][t]{\rotatebox{90}{\tiny\sc IMRM}}%
\psfrag{x31}[t][t]{\rotatebox{90}{\tiny\sc ComN}}%
\psfrag{x32}[t][t]{\rotatebox{90}{\tiny\sc Jac}}%
\psfrag{x33}[t][t]{\rotatebox{90}{\tiny\sc DegP}}%
\psfrag{x34}[t][t]{\rotatebox{90}{\tiny\sc ShP}}%
\psfrag{x35}[t][t]{\rotatebox{90}{\tiny\sc IHW}}%
\psfrag{x36}[t][t]{\rotatebox{90}{\tiny\sc IDB}}%
\psfrag{x37}[t][t]{\rotatebox{90}{\tiny\sc IRM}}%
\psfrag{x38}[t][t]{\rotatebox{90}{\tiny\sc IMHW}}%
\psfrag{x39}[t][t]{\rotatebox{90}{\tiny\sc IMDB}}%
\psfrag{x40}[t][t]{\rotatebox{90}{\tiny\sc IMRM}}%
\psfrag{x41}[t][t]{\rotatebox{90}{\tiny\sc ComN}}%
\psfrag{x42}[t][t]{\rotatebox{90}{\tiny\sc Jac}}%
\psfrag{x43}[t][t]{\rotatebox{90}{\tiny\sc DegP}}%
\psfrag{x44}[t][t]{\rotatebox{90}{\tiny\sc ShP}}%
\psfrag{x45}[t][t]{\rotatebox{90}{\tiny\sc IHW}}%
\psfrag{x46}[t][t]{\rotatebox{90}{\tiny\sc IDB}}%
\psfrag{x47}[t][t]{\rotatebox{90}{\tiny\sc IRM}}%
\psfrag{x48}[t][t]{\rotatebox{90}{\tiny\sc IMHW}}%
\psfrag{x49}[t][t]{\rotatebox{90}{\tiny\sc IMDB}}%
\psfrag{x50}[t][t]{\rotatebox{90}{\tiny\sc IMRM}}%
%
\psfrag{v01}[r][r]{0.4}%
\psfrag{v02}[r][r]{0.6}%
\psfrag{v03}[r][r]{0.8}%
\psfrag{v04}[r][r]{1}%
%
\resizebox{12cm}{!}{\includegraphics{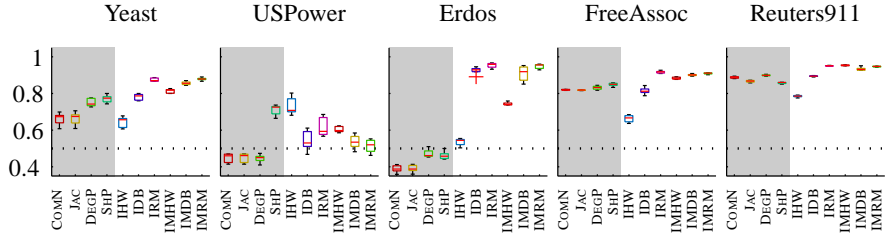}}%
\end{psfrags}%
%
\\
  \vspace{5pt}
  \caption{\textsc{auc} scores for the analysis of the five real networks.}
  \label{fig:RealResults}
\end{figure*}

While the \textsc{imrm} model explicitly accounts for multiple
memberships, the \textsc{irm} model can also implicitly account for
multiple memberships through the between class interactions. To
illustrate this, we analyzed the generated \textsc{hw} and
\textsc{mhw} data by the \textsc{irm} model as well as the proposed
\textsc{imrm} model (see Figure~\ref{fig:SynthExample}). When there
are only single memberships, the \textsc{imrm} reduces to the
\textsc{irm} model; however, when the network is generated such that
the vertices have multiple memberships the \textsc{imrm} model
correctly identifies the ($2\cdot 5=10$) underlying classes. The
\textsc{irm} model on the other hand extracts a larger number of classes
corresponding to all possible ($5^2=25$) combinations of classes
present in the data.  The estimated $\boldsymbol{\rho}$ indicates how
these $25$ classes combine to form the $10$ underlying multiple
membership groups in the network. As such, the \textsc{irm} model has
the same expressive power as the proposed multiple membership models
but interpreting the results can be difficult when multiple membership
community structure is split into several classes with complex
patterns of interaction.

Figure~\ref{fig:SynthResults} shows the link-prediction
\textsc{auc} scores from the analysis of the six generated
networks. Results show that all models work well on data generated
according to their own model or models which they generalize. We
also notice, that the \textsc{irm} model accounts well for multiple
membership structure as discussed and illustrated in
Figure~\ref{fig:SynthExample}. The \textsc{hw} and \textsc{db} models
on the other hand fail in modeling networks with multiple memberships.

\textbf{Real Networks: } We finally analyzed five benchmark complex networks summarized in
Table~\ref{tab:RealNetworkSummary}. The sizes of most of the networks
makes it computationally infeasible for us to analyze them using the
existing multiple-membership approaches proposed in
\cite{Airoldi2008,Meeds2007,Miller2009}. For all the networks,
multiple memberships are conceivable: In protein interaction networks
such as the Yeast network proteins can be part of multiple functional
groups, in social networks such as Erdos scientist
collaborate with different groups of people depending on the research
topic, and in word relation networks such as Reuters911 and FreeAssoc
words can have multiple meanings/contexts. For all these
networks explicitly modeling these multiple contexts can potentially
improve on the structure identification over the equivalent single
membership models.
\begin{table}[tb]
  \caption{Summary of the analyzed real networks: $r$ denotes the networks
    assortativity, $c$ the clustering
    coefficient \cite{Watts1998}, $L$ the average shortest path.}
  \label{tab:RealNetworkSummary}
  \vskip 0.15in
  \begin{center}
    \begin{scriptsize}
      \begin{tabular}{lrrrrrl}
        \hline
        \textsc{Network} & $N$ & $|\mathcal{Y}_1|$ & $r$ & $c$ & $L$& Description\\
        \hline
        Yeast&         2,284&   6,646& -0.10& 0.13&  4.4 & Protein-protein interaction network \dci{Sun2003}\\
        USPower&       4,941&   6,594&  0.00& 0.08& 19.9 & Topology of power grid \dci{Watts1998}\\
        Erdos&         5,534&   8,472& -0.04& 0.08&  3.9 & Erd\"{o}s 02 collaboration network \dci{PajekRep}\\
        FreeAssoc&    10,299&  61,677& -0.07& 0.12&  3.9 & Word relations in free association \dci{FA}\\
        Reuters911&   13,314& 148,038& -0.11& 0.37&  3.1 & Word co-occurence \dci{Corman2002}\\
        \hline
      \end{tabular}
    \end{scriptsize}
  \end{center}
  \vskip -0.1in
\end{table}

\begin{table}\caption{\normalsize  \textbf{Top table:} The number of extracted components for the \textsc{irm} and \textsc{imrm} models. Bold denotes that the number of components are significantly different between the two models (i.e. difference in mean is at least two standard deviations apart). \textbf{Bottom table:} cpu-time usage in hours for 2500  \textsc{irm} and \textsc{imrm} sampling iterations. }
\scriptsize
\centering
\begin{tabular}{c}
\begin{tabular}{l|c|c|c|c|c|}
   & \textbf{Yeast} & \textbf{USPower} & \textbf{Erdos}& \textbf{FreeAssoc} & \textbf{Reuters911} \\
       \hline
      \hline
  \textbf{IRM} & $24.0\pm 0.8$           & $8.6\pm0.4$           & $10.4\pm 0.3$          & $58.6\pm 0.7$    &  $39.8\pm 2.1$      \\
  \hline
  \textbf{IMRM} & $\mathbf{15.4\pm 0.9}$ & $\mathbf{6.8\pm 0.5}$ & $\mathbf{6.8\pm 0.6}$   &$\mathbf{15.6\pm 0.9}$ & $44.8\pm 1.0$   \\
  \hline
\end{tabular}\label{tab:NOC}
\\
\\
\begin{tabular}{l|c|c|c|c|c|}
   & \textbf{Yeast} & \textbf{USPower} & \textbf{Erdos}& \textbf{FreeAssoc} & \textbf{Reuters911} \\
       \hline
      \hline
\textbf{IRM} & $2.3\pm 0.1$           & $4.0\pm 0.2$           & $14.6\pm 5.9$          & $30.1\pm 0.6$    &  $32.5\pm 5.4$      \\
  \hline
  \textbf{IMRM} & $1.7\pm 0.1$ & $8.9\pm0.8$ & $7.1\pm 0.5$   &$28.1\pm 1.9$ & $71.5\pm 3.2$   \\
  \hline
  \end{tabular}
\end{tabular}\label{tab:CPUtime}
\end{table}

In Figure~\ref{fig:RealResults} the \textsc{auc} link prediction score
is given for the five networks analyzed. As can be seen from the
results, modeling multiple memberships significantly improves on
predicting links in the network. In particular when considering the
\textsc{ihw} and \textsc{idb} models and the corresponding proposed
multiple membership models, the learning of structure is improved
substantially for all networks except USPower. Furthermore, it can be
seen that the \textsc{irm} model that can also implicitly account for
multiple memberships in general has a similar performance to the
multiple membership models. The poor identification of structure in the
USPower network might be due to the fact that the average path between
vertices are very high rendering it difficult to detect the underlying
structure for any but the most simple \textsc{ihw} model. While the \textsc{irm} and \textsc{imrm} perform equally well in terms of link prediction it can be seen in table \ref{tab:NOC} that the average number of extracted components for the \textsc{imrm} model is significantly smaller than the number of components extracted by the \textsc{irm} model for all networks except the Reuters911 network where no significant difference is found. As a result, the \textsc{imrm} model is in general able to extract a more compact representation of the latent structure of networks. In table \ref{tab:CPUtime} is given the total cpu-time for estimating the 2500 samples for each of the network using the \textsc{irm} and \textsc{imrm} showing that the order of magnitude for the computational cost of the two models are the same.

\section{Discussion}
While single membership models based on the \textsc{irm} indirectly
can account for multiple memberships as we have shown, the benefit of
the proposed framework is that it allows for these multiple
memberships to be modeled explicitly rather than through complex
between-group interactions based on a multitude of single membership
components. On synthetic and real data we demonstrated that explicitly modeling multiple-membership resulted in a more compact representation of the inherent structure in networks. We further demonstrated that
models that can capture multiple memberships (which includes the
\textsc{irm} model) significantly improve on the link prediction
relative to models that can only account for single membership
structure, i.e., the \textsc{ihw} and \textsc{idb} models. We presently considered undirected networks but we note that the
proposed approach readily generalizes to directed and bi-partite
graphs. Furthermore, the approach also extends to include side
information as proposed in \cite{Miller2009} as well as simultaneous modeling of vertex attributes \cite{Xu2006}. We note however, that
the inclusion of side information requires a linearly scalable
parameterization in order for the overall model to remain
computationally efficient. An attractive property of the \textsc{irm} model over the \textsc{imrm} model is that the \textsc{irm} model admits
the use of collapsed Gibbs sampling which we have found to be more
efficient relative to sampling the non-conjugate multiple membership
models where additional sampling of the $\m{\rho}$ parameter is
required. In future research, we envision combining the \textsc{irm}
and \textsc{imrm} model, using the \textsc{irm} as initialization for
the \textsc{imrm} or by forming hybrid models.

\scriptsize
\itemsep0pt

\end{document}